\documentclass[preprint,prd,amsmath,amssymb,amsthm,nofootinbib]{revtex4-1}
\usepackage{amsmath}
\usepackage{xcolor}
\usepackage{multirow}
\usepackage{array}
\usepackage{booktabs}
\usepackage{threeparttable}

\usepackage[mathscr]{euscript}
\usepackage{bm}
\usepackage{graphicx}
\usepackage{subfigure}
\usepackage{bm}
\usepackage{amsmath,amssymb,amsthm}
\usepackage[colorlinks=true,linkcolor=red]{hyperref}

\usepackage[normalem]{ulem}

\newcommand{\bea}{\begin{eqnarray}}
\newcommand{\eea}{\end{eqnarray}}
\newcommand{\beq}{\begin{equation}}
\newcommand{\eeq}{\end{equation}}

\def\/{\over}

\begin{document}
\title{Entanglement dynamics for two-level quantum systems coupled with massive scalar fields}

\author{Yuebing Zhou$^{1,2}$,
Jiawei Hu$^{1}$\footnote{Corresponding author: jwhu@hunnu.edu.cn} and
Hongwei Yu$^{1}$\footnote{Corresponding author: hwyu@hunnu.edu.cn }}

\affiliation{
$^{1}$Department of Physics and Synergetic Innovation Center for Quantum Effects and Applications, Hunan Normal University, Changsha, Hunan 410081, China\\
$^2$Department of Physics, Huaihua University,  Huaihua 418008, China}

\begin{abstract}

Entanglement is essential in quantum information science. Typically, the inevitable coupling between quantum systems and environment inhibits entanglement from being created between long-distance subsystems and being maintained for a long time. In this paper, we show that when the environment is composed of a bath of massive scalar fields, the region of the separation within which entanglement can be generated is significantly enlarged, and the decay rate of entanglement is significantly slowed down compared with those in the massless case, when the mass of the field $m$ is smaller than but close to the transition frequency of the qubits $\omega$. When $m\geq\omega$, the initial entanglement can be maintained for an arbitrarily long time, regardless of the environmental temperature. Therefore, in principle, it is possible to achieve long-distance entanglement generation and long-lived entanglement by manipulating the energy level spacing of the two-level systems with respect to the mass of the field.

\end{abstract}

\maketitle

\section{Introduction}

Entanglement, as one of the most intrinsic characteristic properties that distinguishes quantum realms from classical realms, was introduced by Schr\"{o}dinger in the early days of the establishment of quantum mechanics \cite{sch}, and has become the core  of quantum information science \cite{information1,information2}.
However, quantum systems are inevitably influenced by the environment they  coupled to, which will lead to  decoherence \cite{decoherence}.
Further studies indicate that, unlike the asymptotic  decoherence of a single quantum system, an initially entangled quantum system composed of two two-level subsystems may get completely disentangled within a finite time even in a vacuum, which is called entanglement sudden death \cite{T. Yu3,  T. Yu},  and has been verified experimentally \cite{esd-3}.
On the other hand, two separable two-level  systems may also get entangled  because the common bath they immersed in can serve as a medium to provide indirect interactions  \cite{Braun,Kim,Schneider,Basharov,Jakobczyk,Reznik,Piani,Z. Ficek, R. Tanas,esb-2,esb-3,esb-4}.
However, entanglement generation  only occurs when the environmental temperature is  low, and the distance between the two subsystems is of the order of or smaller than the transition wavelength, while entanglement sudden death is a general law \cite{esb-1}. Therefore, the possibility of a steady entanglement mediated by an environment is of particular interest and has been extensively studied \cite{Benatti-job,Braun,Liu2007,Chou2008,Benatti2006,Horhammer2008,Paz2008,Shahmoon2013,Tudela2011,Hsiang15}. It has been shown that, in the limit of a vanishing  separation, entanglement can persist in the asymptotic equilibrium state \cite{Benatti-job,Liu2007,Benatti2006,Horhammer2008,Paz2008}. When there is a finite separation between the two subsystems, in order to achieve long-lived entanglement, a cavity/waveguide \cite{Braun,Shahmoon2013,Tudela2011}, or direct coupling between the two subsystems \cite{Hsiang15} is required. Then, a question arises naturally as to whether it is possible to achieve long-distance entanglement generation and long-lived entanglement between two two-level systems in a free space without a direct interaction.  This is what we are going to pursue in the present paper.

We plan to study the entanglement dynamics of a quantum system composed of a pair of two-level subsystems without direct interaction coupled with a massive scalar field in a free space.  
Here it is worth pointing out that the model of two-level atoms interacting with the massless scalar field is frequently used, for simplicity but without loss of generality,  in many studies concerning the atom-field interactions to reveal the basics features concerned, although  a realistic model is multilevel atoms in interaction with  electromagnetic fields in which  the massless scalar fields can be used to  represent their polarization components.  Similarly, a massive scalar field can be used to represent a polarization of a massive vector field.
Our interest in quantum systems interacting with massive fields is twofold.  First, there are elementary quantum fields that are massive, e.g., massive bosons that mediate electroweak interactions in the Standard Model. 
As an explicit example, our model is applicable to neutrinos with two flavors as two-level systems  in interaction with massive boson fields. At this point, let us note that the dissipative effects on neutrino oscillations and the mixed state geometric phase for neutrino oscillations have been investigated in Refs.  \cite{neutrino1,neutrino2,neutrino3,neutrino4}. 
Also, there has been longstanding interest in the possibility of a nonzero photon mass for which massive scalar fields can serve as components of the massive photon field  \cite{mass1,mass2,mass3}, and its implications on the Casimir and Casimir-Polder effects have been studied \cite{Barton1984,Barton1985,Mattioli2019}. 
Second, in certain cases, an environment  can be effectively described by a massive field.  For example, although the  propagation of photons in a waveguide or in a plasma can be perfectly described by the electromagnetic field theory \cite{Shahmoon2013,Tudela2011}, 
it can also be alternatively viewed as photons with an effective mass related to the  critical frequency of the waveguide or the electron plasma density \cite{waveguide,plasma}. More recently, it has been proposed in Ref. \cite{bec} that a two-species BEC can be used to  simulate  massive  scalar field embedded in a curved spacetime. 
Therefore, our study may offer an alternative view on the dynamics of an elementary quantum system in a waveguide or in a BEC, and 
 may shed some light on the control of entanglement generation and protection in an engineered environment.



In a previous work, we have shown that the mass of the field will bring a gray factor to the transition rate of the quantum system \cite{Y. B. Zhou}. That is, for quantum systems coupled with a massive scalar field,  transition rate among different eigenstates will be slowed down compared with that in the massless case. Since the entanglement dynamics of an open quantum system is closely related to the  spontaneous emission of its subsystems, it is of interest to study the entanglement dynamics for a pair of two-level systems coupled with a massive scalar field.  First, we consider an ideal case when the   massive scalar field is  in the Minkowski vacuum state, and then the case when the field is in a thermal state at a finite temperature, which is more   realistic and  where the dissipation effect is stronger.  As we will show in details, even in the thermal bath case,  it is possible to generate entanglement for long-distance subsystems and maintain it for a long time.

\section{The basic formalism}
We consider an open quantum system consisting of two two-level subsystems, interacting with a massive scalar field in the Minkowski vacuum or a thermal bath. The Hamiltonian of the total system takes the form
\begin{equation}
H=H_{A}+H_{F}+H_{I}.
\end{equation}
Here  $H_{A}$ denotes the Hamiltonian of the two two-level systems, which can be written as
\begin{equation}
H_{A}=\frac{\omega}{2}\sigma_{3}^{(1)}+\frac{\omega}{2}\sigma_{3}^{(2)},
\end{equation}
where $\sigma_{i}^{(1)}=\sigma_{i}\otimes\sigma_{0}$, $\sigma_{i}^{(2)}=\sigma_{0}\otimes\sigma_{i}$, with $\sigma_{i}\ (i=1,2,3)$ being the Pauli matrices, $\sigma_{0}$ is the $2\times2$ unit matrix, and $\omega$ is the energy level spacing of the two-level systems. $H_{F}$ is the Hamiltonian of the massive scalar field. The interaction Hamiltonian $H_{I}$, which is supposed to be weak, is taken in analogy to the electric dipole interaction as \cite {J. Audretsch}
\begin{equation}
H_{I}=\mu[\sigma_2^{(1)}\phi(t,\mathbf{x}_1)+\sigma_2^{(2)}\phi(t,\mathbf{x}_2)],
\end{equation}
where $\mu$ is the coupling constant, and $\phi(t,\mathbf{x})$ is the field operator.

To obtain the master equation describing the evolution of the reduced density matrix of the quantum system, we assume that initially the  quantum system is   decoupled from the quantum fields, i.e. the initial state of the total system can be written as $\rho_{\rm tot}(0)=\rho(0)\otimes\rho_{F}$,
where $\rho(0)$  denotes the initial state of the  quantum system, and $\rho_{F}$ is the  state of the massive scalar field. The density matrix of the total system  satisfies the Liouville equation
\begin{equation}
\frac{\partial\rho_{\rm tot}(\tau)}{\partial \tau} = -i[H,\rho_{\rm tot}(\tau)].
\end{equation}
Under the Born-Markov approximation, the reduced density matrix  $\rho(\tau)=\mathrm{Tr}_{F}[\rho_{\rm tot}(\tau)]$ satisfies the Gorini-Kossakowski-Lindblad-Sudarshan (GKLS) master equation \cite{Kossakowski, Lindblad},
\begin{equation}\label{m}
\frac{\partial\rho(\tau)}{\partial \tau} = -i[H_{\rm eff},\rho(\tau)] +\mathcal{L}[\rho(\tau)],
\end{equation}
where
\begin{equation}
H_{\rm eff}=H_{A}-\frac{i}{2}\sum\limits_{\alpha,\varrho=1}^{2}\sum\limits_{i,j=1}^{3}  H_{ij}^{(\alpha\varrho)}\sigma_{i}^{(\alpha)}\sigma_{j}^{(\varrho)},
\end{equation}
and
\begin{equation}
\mathcal{L}[\rho(\tau)]=\frac{1}{2}\sum\limits_{\alpha,\varrho=1}^{2}\sum\limits_{i,j=1}^{3}  C_{ij}^{(\alpha\varrho)}[2\sigma_{j}^{(\varrho)}\rho\sigma_{i}^{(\alpha)}-\sigma_{i}^{(\alpha)}
\sigma_{j}^{(\varrho)}\rho-\rho\sigma_{i}^{(\alpha)}\sigma_{j}^{(\varrho)}].
\end{equation}
Here, $C_{ij}^{(\alpha\varrho)}$ and $H_{ij}^{(\alpha\varrho)}$ can be respectively determined by $\mathcal{G}^{(\alpha\varrho)}(\zeta)$ and $\mathcal{K}^{(\alpha\varrho)}(\zeta)$, which are separately defined by the Fourier and Hilbert transforms of the correlation functions
\begin{equation}\label{G+}
G^{(\alpha\varrho)}(\tau-\tau')=\langle{\phi(\tau,\mathbf{x}_{\alpha})\phi(\tau',\mathbf{x}_{\varrho})}\rangle.
\end{equation}
Thus, $\mathcal{G}^{(\alpha\varrho)}(\zeta)$ and $\mathcal{K}^{(\alpha\varrho)}(\zeta)$ can be expressed as
\begin{equation}\label{F1}
\mathcal{G}^{(\alpha\varrho)}(\zeta)=\int^{\infty}_{-\infty}d\Delta \tau e^{i\zeta\Delta \tau}G^{(\alpha\varrho)}(\Delta\tau),
\end{equation}
\begin{equation}\label{F2}
\mathcal{K}^{(\alpha\varrho)}(\zeta)=\frac{P}{\pi i}\int_{-\infty}^{\infty}d\gamma\frac{\mathcal{G}^{(\alpha\varrho)}(\gamma)}{\gamma-\zeta},
\end{equation}
where $\Delta \tau=\tau-\tau'$, and $P$ represents the principal value. Then, the coefficient matrix $C_{ij}^{(\alpha\varrho)}$  can  be written explicitly as
\begin{equation}\label{c}
C_{ij}^{(\alpha\varrho)}=A^{(\alpha\varrho)}\delta_{ij}-iB^{(\alpha\varrho)}\epsilon_{ijk}\delta_{3k}
-A^{(\alpha\varrho)}\delta_{3i}\delta_{3j},
\end{equation}
where
\begin{eqnarray}\label{b1}
A^{(\alpha\varrho)}&=&\frac{\mu^2}{4}[\mathcal{G}^{(\alpha\varrho)}(\omega)+\mathcal{G}^{(\alpha\varrho)}(-\omega)],\nonumber\\
B^{(\alpha\varrho)}&=&\frac{\mu^2}{4}[\mathcal{G}^{(\alpha\varrho)}(\omega)-\mathcal{G}^{(\alpha\varrho)}(-\omega)].
\end{eqnarray}
Similarly, $H_{ij}^{(\alpha\varrho)}$ in the above equations can be derived by replacing $\mathcal{G}^{(\alpha\varrho)}$ with $\mathcal{K}^{(\alpha\varrho)}$.
Let $A_1=A^{(11)}=A^{(22)}$, $B_1=B^{(11)}=B^{(22)}$, $A_2=A^{(12)}=A^{(21)}$, and $B_2=B^{(12)}=B^{(21)}$,  Eq.~(\ref{c}) can be re-represented as
\bea\label{c1}
&&C_{ij}^{(11)}=C_{ij}^{(22)}=A_{1}\delta_{ij}-iB_{1}\epsilon_{ijk}\delta_{3k}-A_{1}\delta_{3i}\delta_{3j},\nonumber\\
&&C_{ij}^{(12)}=C_{ij}^{(21)}=A_{2}\delta_{ij}-iB_{2}\epsilon_{ijk}\delta_{3k}-A_{2}\delta_{3i}\delta_{3j}.
\eea

To investigate the entanglement dynamics, we need to solve the master equation (\ref{m}).
For convenience, we work in the coupled basis
$\{|G \rangle=|00 \rangle,|A \rangle=\frac{1}{\sqrt{2}}(|10 \rangle-|01 \rangle),|S \rangle=\frac{1}{\sqrt{2}}(|10 \rangle+|01 \rangle),|E \rangle=|11 \rangle\}$.
Then a set of equations describing the evolution of the  system, which are decoupled from other matrix elements, can be expressed in the coupled basis as~\cite{Z. Ficek2}
\bea\label{evolution}
&\dot{\rho}_{G}=-4(A_1-B_1)\rho_{G}+2(A_1+B_1-A_2-B_2)\rho_{A}+2(A_1+B_1+A_2+B_2)\rho_{S},\nonumber\\
&\dot{\rho}_{A}=-4(A_1-A_2)\rho_{A}+2(A_1-B_1-A_2+B_2)\rho_{G}+2(A_1+B_1-A_2-B_2)\rho_{E},\nonumber\\
&\dot{\rho}_{S}=-4(A_1+A_2)\rho_{S}+2(A_1-B_1+A_2-B_2)\rho_{G}+2(A_1+B_1+A_2+B_2)\rho_{E},\nonumber\\
&\dot{\rho}_{E}=-4(A_1+B_1)\rho_{E}+2(A_1-B_1-A_2+B_2)\rho_{A}+2(A_1-B_1+A_2-B_2)\rho_{S},\nonumber\\
&\dot{\rho}_{AS}=-4 A_1 \rho_{AS},\quad\dot{\rho}_{SA}=-4 A_1 \rho_{SA},\nonumber\\
&\dot{\rho}_{GE}=-4 A_1 \rho_{GE},\quad\dot{\rho}_{EG}=-4 A_1 \rho_{EG},
\eea
where $\rho_{IJ}=\langle I|\rho|J\rangle$, $I,J\in\{G,E,A,S\}$. Here, for simplicity, we have abbreviated the diagonal terms of the density matrix elements as follows $\rho_{G}=\rho_{GG},\, \rho_{A}=\rho_{AA},\, \rho_{S}=\rho_{SS},\, \rho_{E}=\rho_{EE}$. Since ${\rho}_{G}+{\rho}_{E}+{\rho}_{A}+{\rho}_{S}=1$, only three of the first four equations in Eq. (\ref{evolution}) are independent. It should be noted that there are still 8 matrix elements whose  evolution equations are not given here. It is because that, if the initial density matrix is chosen as  of the X form, i.e. the only nonzero elements are those along the diagonal and anti-diagonal elements in the decoupled basis $\{|00\rangle,|01\rangle,|10\rangle,|11\rangle\}$, and its dynamics can be described with the second order Markovian master equation, then the X form will be maintained during the  evolution \cite{xstate}.

The degree of entanglement can be measured by concurrence \cite{W. K. Wootters} and negativity \cite{Zyczkowski1998,Vidal2002},  which range from 0 for separable states, to 1 for maximally entangled states. For two-qubit pure states, the negativity is the same as the concurrence. However, they differ for mixed states. 
Here, we characterize the degree of entanglement by both of them, and see if the effects of the mass of the field on entanglement are dependent on the entanglement quantifier we choose. 
For the X-type states, the concurrence  takes the form \cite{R. Tanas}
\begin{eqnarray}\label{C}
C[\rho(\tau)]=\mathrm{max}\{0,K_{1}(\tau),K_{2}(\tau)\},
\end{eqnarray}
where
\begin{eqnarray}\label{K1 K2}
&&K_{1}(\tau)=\sqrt{[\rho_{A}(\tau)-\rho_{S}(\tau)]^{2}-[\rho_{AS}(\tau)-\rho_{SA}(\tau)]^{2}}-2\sqrt{\rho_{G}(\tau)\rho_{E}(\tau)},\nonumber\\
&&K_{2}(\tau)=2|\rho_{GE}(\tau)|-\sqrt{[\rho_{A}(\tau)+\rho_{S}(\tau)]^{2}-[\rho_{AS}(\tau)+\rho_{SA}(\tau)]^{2}},
\end{eqnarray}
and the negativity can be expressed as
\begin{eqnarray}\label{N}
\mathcal{N}[\rho(\tau)]=\sum_{i}\mathrm{max}\{0,-2 N_{i}(\tau)\},
\end{eqnarray}
where
\begin{eqnarray}\label{N1 N2}
&&N_{1}(\tau)=\frac{1}{2}\left(\rho_{G}(\tau)+\rho_{E}(\tau)-\sqrt{[\rho_{A}(\tau)-\rho_{S}(\tau)]^{2}-[\rho_{AS}(\tau)-\rho_{SA}(\tau)]^{2}+[\rho_{G}(\tau)-\rho_{E}(\tau)]^{2}}\right),\nonumber\\
&&N_{2}(\tau)=\frac{1}{2}\left(\rho_{A}(\tau)+\rho_{S}(\tau)-\sqrt{4|\rho_{GE}(\tau)|^2+[\rho_{AS}(\tau)+\rho_{SA}(\tau)]^{2}}\right).
\end{eqnarray}
\section{The impact of mass on entanglement evolution}
In this section, we will focus on the effect brought by the mass of the scalar field on  entanglement evolution.
First, we discuss  an ideal case when the   massive scalar field is  in the Minkowski vacuum state,  and then the thermal bath case, which is more realistic and with a stronger  dissipation effect.

\subsection{In the Minkowski  vacuum}
We consider two two-level systems which are separated from each other by a distance $L$ along the $z$ axis. Then the trajectories of the two static two-level  systems can be written as
\begin{eqnarray}\label{trajectories}
&&t_{1}(\tau)=\tau,\ \  x_{1}(\tau)=0,\ \  y_{1}(\tau)=0,\ \  z_{1}(\tau)=0,\nonumber\\
&&t_{2}(\tau)=\tau,\ \  x_{2}(\tau)=0,\ \  y_{2}(\tau)=0,\ \  z_{2}(\tau)=L,
\end{eqnarray}
with $\tau$ being the proper time. Here, for simplicity, we do not take into account the effect of back reaction of the fields on the trajectories of the two-level systems. The Wightman function of the massive scalar field in the Minkowski vacuum can be expressed as
\begin{eqnarray}\label{correlation1}
G^{(\alpha\varrho)}\left(\Delta\tau\right)&=&\frac{1}{4\pi^2}\int_{m}^{\infty}\frac{\sin{\left(\sqrt{\omega_k^2-m^2}|\Delta\vec{x}_{\alpha\varrho}|\right)}}
{|\Delta\vec{x}_{\alpha\varrho}|}\;e^{-i \omega_k\Delta t}\;d\omega_k,
\end{eqnarray}
where the spatial interval $|\Delta\vec{x}_{\alpha\varrho}|=\sqrt{(x_{\alpha}-x_{\varrho})^2+(y_{\alpha}-y_{\varrho})^2+(z_{\alpha}-z_{\varrho})^2}$, and the time interval  $\Delta t=t-t'$. Substituting the trajectories (\ref{trajectories}) into Eq. (\ref{correlation1}), we  get
\begin{eqnarray}\label{correlation2}
&&G^{(11)}(\Delta\tau)=G^{(22)}(\Delta\tau)=\frac{1}{4\pi^2}\int_{m}^{\infty}\sqrt{\omega_k^2-m^2}\;e^{-i \omega_k\Delta \tau}\;d\omega_k,\nonumber\\
&&G^{(12)}(\Delta\tau)=G^{(21)}(\Delta\tau)=\frac{1}{4\pi^2}\int_{m}^{\infty}\frac{\sin{\left(L\sqrt{\omega_k^2-m^2}\right)}}
{L}e^{-i \omega_k\Delta \tau}\;d\omega_k.
\end{eqnarray}
By a substitution of Eq. (\ref{correlation2}) into Eq. (\ref{F1}) and then Eq. (\ref{b1}), one can directly obtain the coefficients $A_{1}$, $B_{1}$ and $A_{2}$, $B_{2}$ in Eq. (\ref{c1}) as
\bea\label{AB1}
&&A_1=B_1=\frac{\Omega\Gamma_0}{4}\;,\;\;\;\;\;\;\;\;\;\;\;\;\;\;\;\;\;\;A_2=B_2=\frac{\Omega\Gamma_0}{4}\lambda.
\eea
Here, $\Gamma_0=\mu^2\omega/{2\pi}$ is the spontaneous emission rate of  quantum systems coupled with massless scalar field in vacuum, the dimensionless factor $\Omega$, which is a function of $m$ and $\omega$, can be written as
\begin{eqnarray}\label{Omega1}
\Omega(m,\omega)=
\begin{cases}
\sqrt{1-\frac{m^2}{\omega^2}},& \omega>m,\\
0,& \omega\leq m,
\end{cases}
\end{eqnarray}
and the factor $\lambda$ defined as $\lambda=A_2/A_1=B_2/B_1$, can be expressed as
\bea\label{lambda}
\lambda=\frac{\sin(\omega L\Omega)}{\omega L\Omega}.
\eea

Plugging Eq. (\ref{AB1}) into Eq. (\ref{evolution}), we can  get the general solutions of the  evolution equations and the general expressions of concurrence and negativity which are given in Appendix \ref{general solution}.  The concurrence and negativity can be   written as a function of $ m, \omega, \rho(0) , L, \tau$,  where $\rho(0)$ is a parameter related to the initial state of the quantum system. From the general solution of the time-dependent density matrix elements Eq.~(\ref{general solution1}), we can easily prove that the concurrence and negativity satisfy the following  relation
\bea\label{solutionC1}
&&C\left[m,\omega,\rho(0),L,\tau\right]=C\left[0,\omega,\rho(0),\Omega L,\Omega\tau\right],\\
&&\mathcal{N}\left[m,\omega,\rho(0),L,\tau\right]=\mathcal{N}\left[0,\omega,\rho(0),\Omega L,\Omega\tau\right].\label{solutionN1}
\eea
It shows that there is a correspondence between the concurrence (and negativity) of quantum systems coupled with massive and massless scalar fields, i.e. a substitution of  $\Omega L$  for  $L$, and  $\Omega\tau$ for $\tau$  respectively. In the  following, we discuss the effects of the mass of  the field in details.

\subsubsection{Long-lived entanglement}

It  is  clear from Eqs. \eqref{solutionC1} and \eqref{solutionN1} that  $\Omega$ plays the role of a time-delay factor since $\Omega<1$. No matter what the initial state is, what type of evolution the system undergoes, and what the entanglement quantifier is, the evolution time $t$ is $1/\Omega$ times that of the massless case $t_0$, i.e.
\bea\label{time-delay1}
t=\frac{t_0}{\Omega}=\frac{t_0}{\sqrt{1-m^2/{\omega^2}}},
\eea
when $m<\omega$. That is, when $m<\omega$, the larger the mass, the slower the entanglement evolution.
Physically,  the time-delay effect can be understood as the decrease of the transition rate  brought up by the mass of the field \cite{Y. B. Zhou}.
When  $m\geq \omega$, the time-delay factor $\Omega=0$, the quantum system will be locked up in the initial state. This is due to the fact that the entanglement generation and degradation is  via the the excitation  and de-excitation   of the two-level systems, which are possible only when $m< \omega$. This is the  selection rule for   quantum systems coupled with a  massive scalar field \cite{Y. B. Zhou}. Therefore, the initial entanglement can be maintained as  long as we want by  manipulating   the energy level spacing of the two-level systems with respect to the mass of the field.


For a concrete example, we take  $ \lambda = \frac{\sin(\omega L\Omega)}{\omega L\Omega} = 0 $, which corresponds to the cases   when the subsystems are infinitely far from each other (in independent baths), or when the  separation $L$ satisfies  $ \Omega \omega L = n \pi $ (where $ n $ is a positive integer),
and the initial state of the system is taken as $\rho_E(0)=e$, $\rho_G(0)=g$, $\rho_A(0)=a$, $\rho_S(0)=s$, and $\rho_{IJ}(0)=0$ when $I\neq J$.
Then, it can be found from the general expression  of  concurrence \eqref{K1(lambda0)} or    negativity \eqref{N1(lambda0)}  that  the  necessary and  sufficient  condition that the system get disentangled within a finite time is $4e g<(a-s)^2<4e$, and  the lifetime of entanglement $ \tau $ is
\bea\label{survival time}
\tau=\frac{1}{\Omega\Gamma_0}\ln\frac{2 e \left(\sqrt{2 (a+e)^2+2 (e+s)^2-4 e}+a+2 e+s\right)}{4 e-(a-s)^2}=\frac{\tau_0}{\Omega}.
\eea
Here, $\tau_0$ is the lifetime in the massless case. This indicates  that  the lifetime of the entanglement for  quantum systems coupled with massive scalar field can be much longer than those coupled with massless scalar field when $\Omega$ is close to 0.


\subsubsection{Long-distance generation of entanglement}

It is well-known that, for quantum systems coupled with massless fields, the two separable subsystems can get  entangled only when the separation is smaller than or of the order of the transition wavelength  \cite{esb-1}. However, it is clear from Eqs. \eqref{solutionC1} and \eqref{solutionN1} that, no matter what the entanglement quantifier is, the range of  the  separation is  $\Omega^{-1}$ that of the massless case,  which can be much larger than the transition wavelength.

\begin{figure}[!htbp]
\centering
\includegraphics[width=0.495\textwidth]{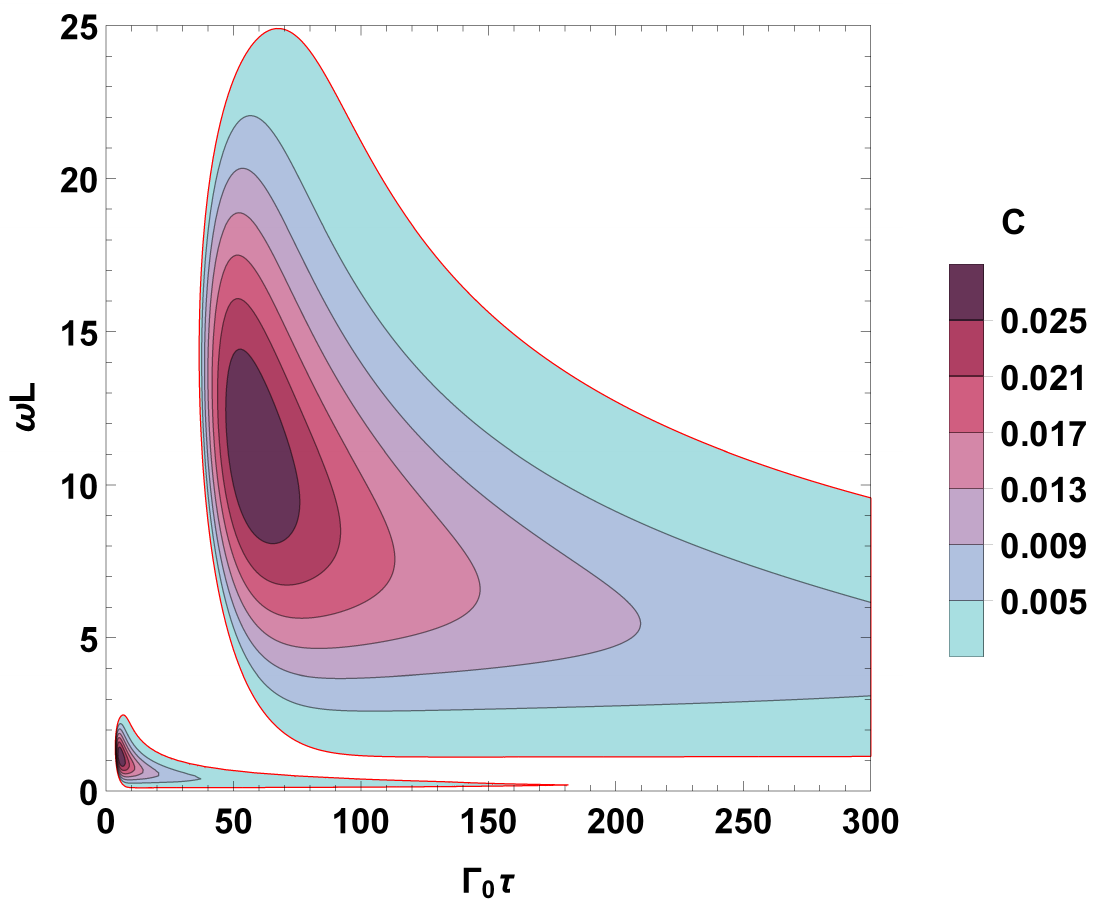}
\includegraphics[width=0.495\textwidth]{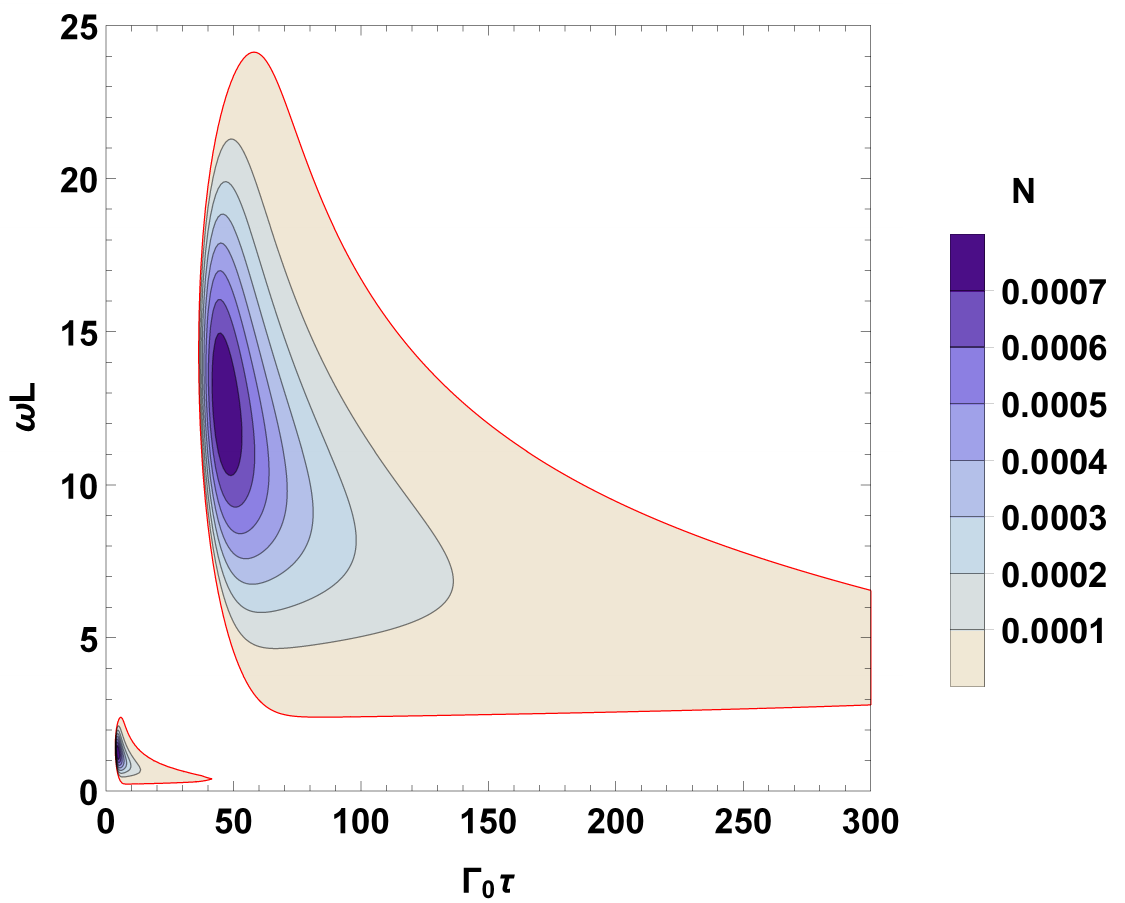}
\caption{\label{C(t,L)m0m0995}
Comparison between the dynamics of entanglement quantified by concurrence (left) and negativity (right) for two two-level quantum systems initially prepared in $|E\rangle$ coupled to massive scalar field with $m/\omega=0.995$ (the larger region)  and massless one  (the smaller region) in the Minkowski vacuum.
Note that  contour lines with $C<10^{-3}$ and $\mathcal{N}<10^{-5}$ are not drawn here.}
\end{figure}



As an example, we assume  the quantum system is initially prepared in $|E\rangle$, and calculate the time evolution of concurrence and negativity for different  separations $\omega L$ as shown in Fig. \ref{C(t,L)m0m0995}. Compared with the massless case, the region of the  separation within which entanglement can be generated is enlarged 10 times when $m/\omega=0.995$. At the same time, the lifetime of entanglement is 10 times that of the massless case. As $m/\omega$ gets closer to 1, the region of the  separation within which entanglement can be generated is larger,  the lifetime of entanglement is longer, and the birth time of entanglement is postponed  at the same time.



\subsection{In a thermal bath}

Interestingly, as shown in Ref. \cite{T. Yu3}, in the vacuum case,  entanglement sudden death can be avoided by choosing an appropriate initial state \cite{T. Yu3}.  However, for  two subsystems with a nonvanishing separation immersed in a thermal bath, entanglement sudden  death is unavoidable~\cite{esb-1}. In this section, we turn to the thermal bath case, and see whether the long-lived entanglement and long-distance generation of entanglement can still be achieved.

First, we work out the Wightman function of the massive scalar field at a finite temperature $T=\beta^{-1}$, which can be written as an infinite imaginary-time image sum of the vacuum Wightman functions as \cite{qftcs}
\begin{eqnarray}\label{wight-relation}
G^{(\alpha\varrho)}_{\beta}\left(\Delta\tau\right)=\sum
\limits_{n=-\infty}^{+\infty}G^{(\alpha\varrho)}\left(\Delta\tau+i n\beta\right).
\end{eqnarray}
Similar to the calculation in the vacuum case, by substituting the vacuum Wightman function (\ref{correlation1}) and the trajectories  (\ref{trajectories}) into Eq. (\ref{wight-relation}), we can get $G^{(\alpha\varrho)}_{\beta}(\Delta\tau)$
by exchanging the order of integration for $\omega_k$ and the sum for $n$.
Then, putting it into Eq. (\ref{F1}) and Eq. (\ref{b1}), one can directly obtain the coefficients $A_{1}$, $B_{1}$ and $A_{2}$, $B_{2}$ in Eq. (\ref{c1}) as
\bea\label{AB}
&&A_1=\frac{\Omega\Gamma_0}{4}\coth{\omega\beta\/2}\;,\;\;\;\;\;A_2=\frac{\Omega\Gamma_0}{4}\lambda\coth{\omega\beta\/2}\;,\nonumber\\
&&B_1=\frac{\Omega\Gamma_0}{4}\;,\;\;\;\;\;\;\;\;\;\;\;\;\;\;\;\;\;\;B_2=\frac{\Omega\Gamma_0}{4}\lambda\;.
\eea
Note that the  factors $\Omega$ and $\lambda$ have been defined  in Eq. (\ref{Omega1}) and Eq. (\ref{lambda}) respectively.

Similarly,
by substituting Eq. \eqref{AB} into Eqs. \eqref{evolution} and \eqref{C}, we can easily prove that the concurrence and negativity satisfy the following  relation
\bea\label{solutionC}
&&C\left[m,\omega,T,\rho(0),L,\tau\right]=C\left[0,\omega,T,\rho(0),\Omega L,\Omega\tau\right],\\
&&\mathcal{N}\left[m,\omega,T,\rho(0),L,\tau\right]=\mathcal{N}\left[0,\omega,T,\rho(0),\Omega L,\Omega\tau\right].
\eea
This shows that for two two-level systems in a thermal bath, the effect of the mass of the field on the entanglement evolution is the same as that in the Minkowski vacuum.

\begin{figure}[!htbp]
\centering
\includegraphics[width=0.495\textwidth]{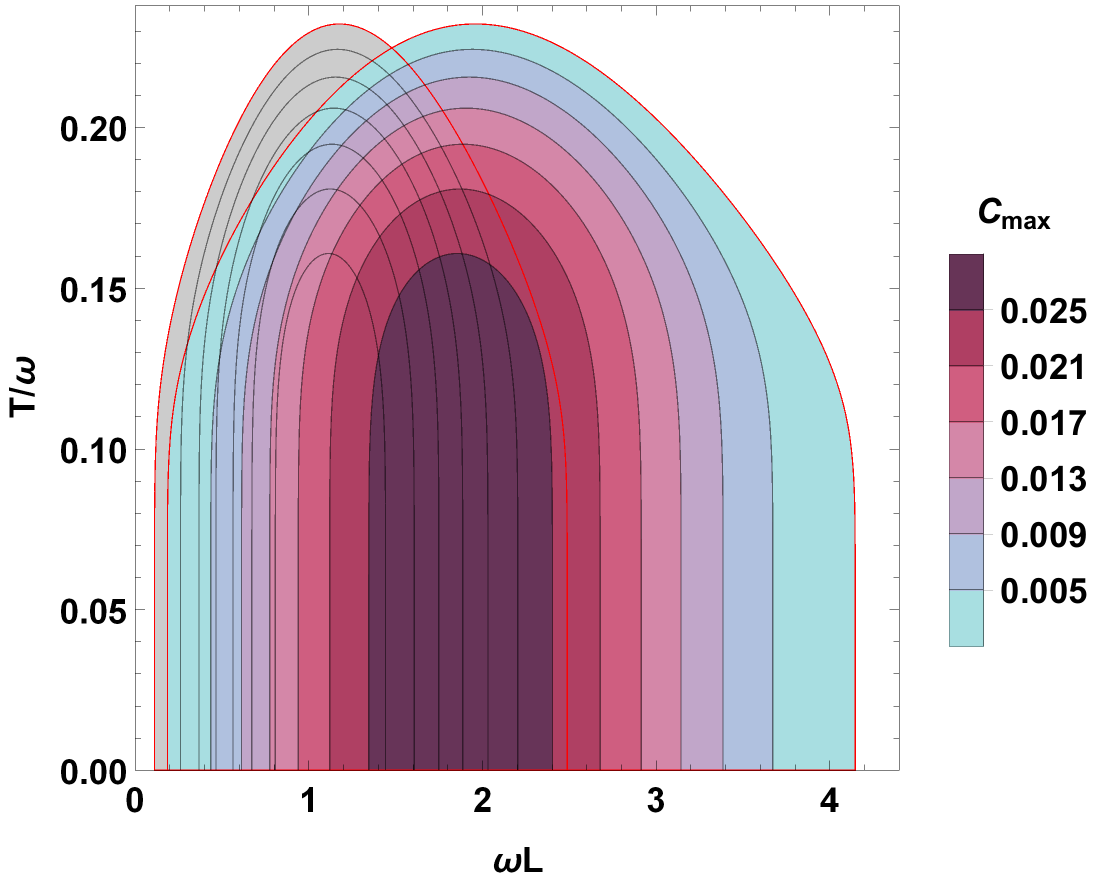}
\includegraphics[width=0.495\textwidth]{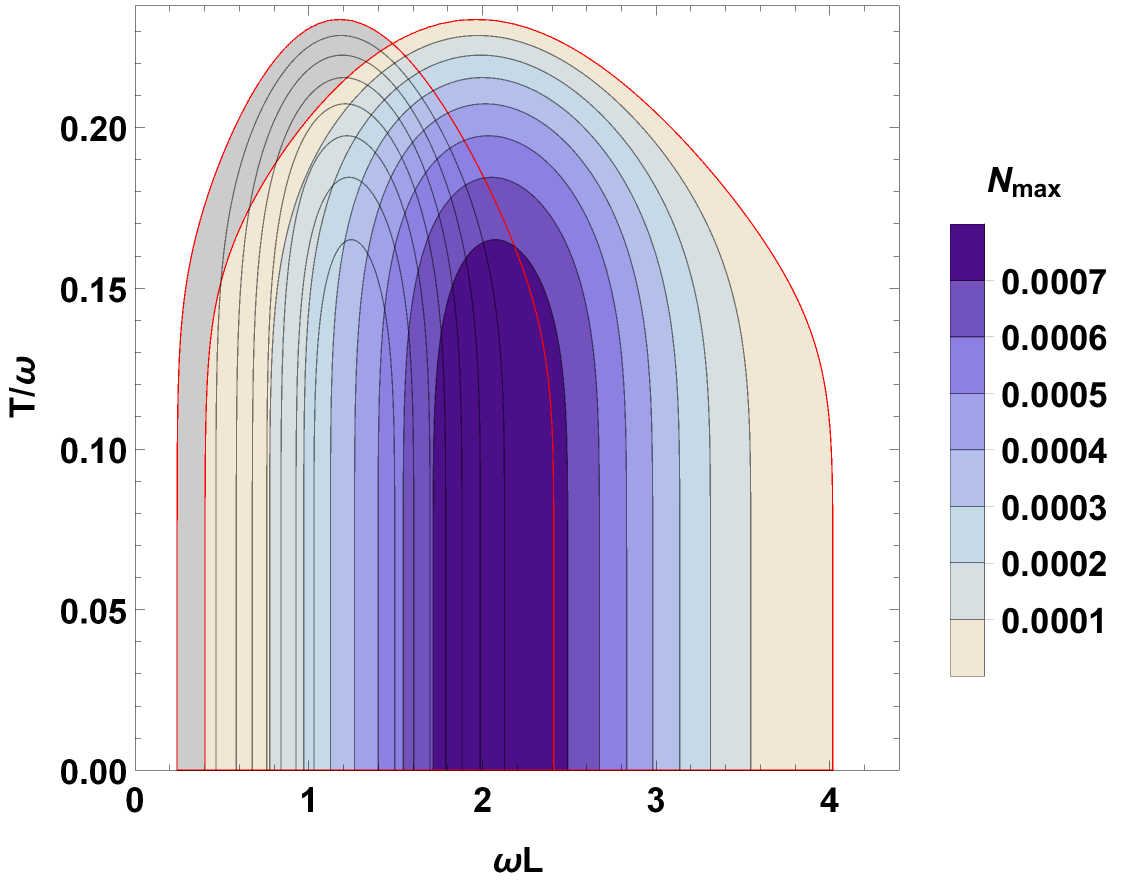}
\caption{\label{tt08}
The  maximum of concurrence (right) and negativity (left) during evolution for two two-level quantum systems initially prepared in $|E\rangle$ coupled to massive scalar field with $m/\omega=0.8$ (the colored region)  and massless one  (the uncolored region)  in the Minkowski vacuum. Note that  contour lines with $C_{max}<10^{-3}$ and $\mathcal{N}_{max}<10^{-5}$ are not drawn here.}
\end{figure}

As an example, in  Fig. \ref{tt08}, we show the range of  the temperature and  separation  within which entanglement can be generated for a quantum system initially prepared in $| E\rangle$. The contour plot gives the maximum of concurrence (left) and negativity (right) in the whole process of  evolution. It is obvious that the range of  the  separation  within which entanglement can be generated has been enlarged compared with that in the massless case, while the temperature range remains the same. When  the thermal wavelength $\beta$ is larger or comparable with the transition wavelength  $\omega^{-1}$, or to be specific when $T/\omega<0.23$, the long-lived entanglement and long-distance generation of entanglement can still be achieved if the  mass of the field is close to but smaller than the transition frequency, i.e., $m\lesssim\omega$, and the created entanglement can be kept for a long time. When $m\geq\omega$, entanglement cannot be generated for initially separable subsystems, but for initially entangled quantum systems, the  entanglement can be kept for an arbitrary long time regardless of the environmental temperature. This is in sharp contrast to the  massless
 field case where  entanglement sudden  death is shown to be unavoidable~\cite{esb-1}.

\section{Summary}

In this paper, we have investigated, in the framework of open quantum systems, the entanglement dynamics for a quantum system composed of a  pair of two-level subsystems coupled with a bath of fluctuating massive scalar fields in the Minkowski vacuum and in a thermal bath at a finite temperature.
We find that  the region of the  separation  of the subsystems within which entanglement can be generated is significantly enlarged, and the decay rate of entanglement is significantly slowed down compared with those in the massless case,  when the mass of the field  $m$ is smaller than but close to the energy level spacing $\omega$.  In particular, when $m\geq \omega$,   the initial entanglement of the quantum system can be maintained for an arbitrarily long time, regardless of the environmental temperature. 
We characterize the degree of entanglement by both concurrence and negativity, and show that  the effects of the mass of the field on entanglement are general, and are not peculiar of the entanglement quantifier we choose. 
Therefore, in principle, it is possible to achieve long-distance entanglement generation and long-lived entanglement  by  manipulating   the energy level spacing with respect to the mass of the field.


\begin{acknowledgments}
This work was supported in part by the NSFC under Grants No. 11805063, No. 11690034, and No. 12075084, and the Hunan Provincial Natural Science Foundation of China under Grant No. 2020JJ3026.
\end{acknowledgments}

\appendix

\section{The analytical expression of the time-dependent concurrence and negativity}\label{general solution}
In this Appendix, we solve the time evolution of the open system analytically and give the expression  of time-dependent concurrence and negativity.
Substituting Eq. (\ref{AB1}) into
Eq. (\ref{evolution}), we can easily get the general solution of the time evolution equations of the  system as
\bea\label{general solution1}
&&\rho_{G}(\tau)=1+\xi^2\;\frac{1+3\lambda^2}{1-\lambda^2}\rho_{E}(0)
-\xi\;[f_{A}(-\lambda,\xi)+f_{S}(\lambda,\xi)],\;\;\;\;\;\;\;\;\rho_{E}(\tau)=\xi^2\rho_{E}(0),\nonumber\\
&&\rho_{A}(\tau)=-\xi^2\;\frac{1-\lambda}{1+\lambda}\rho_{E}(0)+\xi\;f_A(-\lambda,\xi),
\;\;\;\rho_{S}(\tau)=-\xi^2\;\frac{1+\lambda}{1-\lambda}\rho_{E}(0)+
\xi\;f_S(\lambda,\xi),\nonumber\\
&&\rho_{GE}(\tau)=\xi\;\rho_{GE}(0),\;\;\;\;\;\;\;\;\;\;\;\;\;\;\;\rho_{EG}(\tau)=\xi\;\rho_{EG}(0),\nonumber\\
&&\rho_{AS}(\tau)=\xi\;\rho_{AS}(0),\;\;\;\;\;\;\;\;\;\;\;\;\;\;\;\;\rho_{SA}(\tau)=\xi\;\rho_{SA}(0),
\eea
where the function $f(\lambda,\xi)$ can be expressed as
\begin{eqnarray}\label{f}
f_I(\lambda,\xi)=\left(\frac{1+\lambda}{1-\lambda}\rho_{E}(0)+\rho_{I}(0)\right)\xi^{\lambda}.
\end{eqnarray}
Here $I\in\{A,S\}$, and  $\xi$ is an exponential decay function of the proper time $\tau$, i.e.,
\begin{eqnarray}\label{xi}
\xi(\tau)=e^{-\Omega\Gamma_0 \tau}.
\end{eqnarray}
Note that the explicit  forms of $ \Omega $ and $ \lambda $ have been given in Eq. (\ref{Omega1}) and Eq.  (\ref{lambda}), respectively.


When  $\rho_{AS}(0)=\rho_{SA}(0)=0$, substituting Eq. (\ref{general solution1}) into Eq. (\ref{K1 K2}), we can obtain the general forms of the   $K_1(\tau)$ and $K_2(\tau)$ which are related to concurrence, as
\bea\label{general K1}
&&K_1(\tau)=\xi\left|\xi\frac{4\lambda}{1-\lambda^2}\rho_{E}(0)+g(\lambda,\xi)\right|
-2\xi\sqrt{\xi^2\;\frac{1+3\lambda^2}{1-\lambda^2}\rho_E^2(0)+[1-\xi\;h(\lambda,\xi)]\rho_E(0)},\nonumber\\
&&K_2(\tau)=\xi\left(2|\rho_{GE}(0)|+2\xi\frac{1+\lambda^2}{1-\lambda^2}\rho_{E}(0)-h(\lambda,\xi)\right),\label{general K2}
\eea
and the  $N_1(\tau)$ and $N_2(\tau)$ which are related to negativity, as
\bea\label{general N1}
N_1(\tau)&=&\xi^2\;\frac{1+\lambda^2}{1-\lambda^2}\rho_E(0)+\frac{1-\xi\;h(\lambda,\xi)}{2}\nonumber\\
&&-\frac{1}{2}\sqrt{\left(\xi^2\frac{4\lambda}{1-\lambda^2}\rho_{E}(0)+\xi g(\lambda,\xi)\right)^2+\left(\xi^2\;\frac{4\lambda^2}{1-\lambda^2}\rho_E(0)+1-\xi\;h(\lambda,\xi)\right)^2},\nonumber\\
N_2(\tau)&=&\frac{\xi}{2}\left(h(\lambda,\xi)-2\xi\frac{1+\lambda^2}{1-\lambda^2}\rho_{E}(0)-2|\rho_{GE}(0)|\right),\label{general K2}
\eea
where the functions of $g(\lambda,\xi)$ and $h(\lambda,\xi)$ are defined with  $f_I(\lambda,\xi)$ as follows
\begin{eqnarray}\label{gh}
&g(\lambda,\xi)=f_{A}(-\lambda,\xi)-f_{S}(\lambda,\xi),\nonumber\\
&h(\lambda,\xi)=f_{A}(-\lambda,\xi)+f_{S}(\lambda,\xi).
\end{eqnarray}
If $\lambda=0$, then $g(\lambda,\xi)=\rho_A(0)-\rho_S(0)$, and $h(\lambda,\xi)=1+\rho_E(0)-\rho_G(0)$, the concurrence coefficients $K_1(\tau)$ and $K_2(\tau)$ can be simplified as
\bea\label{K1(lambda0)}
&&K_1(\tau)=\xi\left|\rho_A(0)-\rho_S(0)\right|
-2\xi\sqrt{\xi^2\;\rho_E^2(0)-\xi\;[1+\rho_E(0)-\rho_G(0)]\rho_E(0)+\rho_E(0)},\nonumber\\
&&K_2(\tau)=\xi\big[2|\rho_{GE}(0)|+(2\xi-1)\rho_{E}(0)-1+\rho_G(0)\big],
\eea
and the negativity coefficients $N_1(\tau)$ and $N_2(\tau)$ can be simplified as
\bea\label{N1(lambda0)}
N_1(\tau)&=&\xi^2\;\rho_E(0)+\frac{1}{2}(1-\xi)-\frac{1}{2}\xi\left[\rho_E(0)-\rho_G(0)\right]\nonumber\\
&&-\frac{1}{2}\sqrt{\xi^2 \left[\rho_A(0)-\rho_S(0)\right]^2+\left[4\xi^2\rho_E(0)+1-\xi-\xi\big(\rho_E(0)-\rho_G(0)\big)\right]^2},\nonumber\\
N_2(\tau)&=&\frac{\xi}{2}\left[1-\rho_G(0)-(2\xi-1)\rho_{E}(0)-2|\rho_{GE}(0)|\right].\label{general K2}
\eea

\end{document}